\documentclass[conference,a4paper]{IEEEtran}

\usepackage{xcolor}
\usepackage{balance}
\usepackage{cite}

\ifCLASSINFOpdf
  \usepackage[pdftex]{graphicx}
\else
  \usepackage[dvipdfmx]{graphicx}
\fi

\usepackage{amsmath}

\ifCLASSOPTIONcompsoc
 \usepackage[caption=false,font=normalsize,labelfont=sf,textfont=sf]{subfig}
\else
 \usepackage[caption=false,font=footnotesize]{subfig}
\fi

\graphicspath{
{./_figs/}
{/Users/kyamamot/Dropbox/due/240804_pico/_out/240914_123715/figs/}
}

\renewcommand{\baselinestretch}{0.97}

\usepackage{amsfonts}
\usepackage{datetime}
\usepackage{booktabs}
\usepackage{mathtools}
\def\jj{{\rm j}}
\def\ee{{\rm e}}

\DeclareMathOperator*{\argfirstpeak}{arg\,FirstSignal}

\begin{document}
%
\title{Delay-Synchronous Wideband Channel Sounding Using Off-The-Shelf Multi-Antenna WiFi Devices}

\author{\IEEEauthorblockN{
Koji Yamamoto\IEEEauthorrefmark{1} and   
Katsuyuki Haneda\IEEEauthorrefmark{2}   
}                                     
\IEEEauthorblockA{\IEEEauthorrefmark{1}
Faculty of Information and Human Sciences,
Kyoto Institute of Technology,
Kyoto, Japan, kyamamot@kit.ac.jp}
\IEEEauthorblockA{\IEEEauthorrefmark{2}
Department of Electronics and Nanoengineering, Aalto University, Espoo, Finland, katsuyuki.haneda@aalto.fi}
}



\maketitle


\begin{abstract}

It has been shown that WiFi devices enable sensing of environments and targets through their channel state information. However, the same devices have not been used for delay-synchronous channel sounding due to challenges related to the stability of synchronization and lack of reference power levels. Due to factors such as uncertainty in symbol reception timing, impulse responses are discontinuous across acquisitions. The present paper addresses the challenges to perform delay-synchronous channel sounding using off-the-shelf multiple-antenna IEEE 802.11ax WiFi devices, referred to as SoundiFi.
Stable delay synchronization and power level reference are realized by remoting the antennas with coaxial cables and devoting one of the antennas as a reference channel, with which the gain and delay of other simultaneous channels are defined. Indoor experiments confirmed that the impulse response becomes continuous across successive acquisitions and provide the absolute delay. The impulse response has a noise level at $-\text{115\,dB}$, indicating the maximum path gain value that can be measured with the devices. The impulse response also revealed the existence of long-delayed multipaths up to 132\,m propagation distance in a reverberant 30-m-long corridor.

\end{abstract}

\vskip0.5\baselineskip
\begin{IEEEkeywords}
 delay-synchronous channel sounding, channel state information, IEEE 802.11ax, WiFi sensing.
\end{IEEEkeywords}

%

\section{Introduction}
\label{sec:introduction}

There has been remarkable progress in WiFi sensing using channel state information (CSI) \cite{Ma2019wifi}, along with the development of CSI acquisition tools.
For example, by running PicoScenes \cite{Jiang2022IOT} on a commodity off-the-shelf (COTS) device equipped with an Intel AX210 WiFi card, it is possible to acquire $2\times 2$ multiple-input-multiple-output orthogonal frequency division multiplexing (MIMO-OFDM) CSI in IEEE 802.11ax \cite{ieee80211ax}, with a 160 MHz bandwidth and 2025 subcarriers.
These values are distinct from those used in early WiFi sensing based on IEEE 802.11n, which utilized a bandwidth of up to 40\,MHz and only several tens of subcarriers \cite{Halperin2011SIGCOMM}.

CSI acquisition is a form of channel sounding; however, achieving \textit{delay-synchronous} channel sounding with WiFi COTS devices, which results in a continuous impulse response in terms of delay across acquisitions, is challenging.
For channel sounders, ``correct synchronization between Tx and Rx is especially important'' \cite[\S8]{Molisch2011}, but WiFi operates differently.
The challenge is primarily due to the use of OFDM, where its symbol reception timing has a degree of freedom within the cyclic prefix.
Moreover, the stability of internal reference clocks in transceiver devices significantly impacts synchronization.
Consequently, the delay of the impulse response obtained from the CSI varies over acquisitions.

The lack of delay-synchronization in CSI measurements imposes various constraints on WiFi sensing techniques.
A typical example is that, while the acquired CSI is a complex sequence that inherently contains delay-related information in its phase, e.g., a target to detect for security purposes and for improved communications \cite{Tamai2024TVT,Yasuba2023GLOBECOM}, only the magnitude is used as input due to the non-delay-synchronous nature.

In this paper, we propose and experimentally demonstrate the feasibility of delay-synchronous channel sounding framework, termed \textit{SoundiFi}, using COTS IEEE 802.11ax WiFi devices, which are capable of $2 \times 2$ MIMO data transfer with up to 160\,MHz bandwidth.
Achieving delay-synchronous channel sounding allows for an easier realization of the existing sensing or novel sensing techniques such as tracking targets of interests in dynamic environments. 

The primary contributions of this study are summarized as follows:
(i) We propose a delay-synchronous channel sounding framework based on COTS WiFi devices, termed SoundiFi.
       The key aspects of SoundiFi are remoting the receiver (Rx) antennas using coaxial cables and using it as a reference antenna with a known free-space propagation distance near the Tx, enabling the calibration of the propagation delay between the two communicating devices.
(ii) After the proposed delay calibration, we confirm that the impulse response derived from the observed CSI reveals the absolute delay of radio channels and remains continuous across observations in dynamic environments.
      Furthermore, from the observations conducted in a 30-m-long corridor, we successfully detected delayed paths with propagation distances of up to 132\,m.

In this study, although we utilize devices and firmware for WiFi sensing, the use of antenna remoting makes it challenging to directly apply SoundiFi to typical WiFi sensing scenarios using everyday WiFi devices.
Instead, this study demonstrates that delay-synchronous channel sounding can serve as a novel application for devices and firmware enabling WiFi sensing, beyond their previously considered use cases.


\section{Impact of Delay-Synchronous Channel Sounding Using COTS WiFi Devices}

\subsection{Conventional Delay-Synchronous Channel Sounding}

SoundiFi makes delay-synchronous channel sounding---previously difficult to conduct---accessible to a wider range of researchers than those relying on instrument-grade transceiver devices and clocks.
This provides a cost-effective and license-exempt alternative to conventional short-range channel sounding.

From a license-exempt perspective, since only WiFi devices are used for transmission and reception, unlike the aforementioned conventional channel sounding methods, it is easier to implement in terms of radio regulations.
Although coaxial cables were inserted between the antenna and the Rx, no modifications were made to the transmitter (Tx).
Another advantage of using WiFi devices, from a different perspective, is that SoundiFi is inherently resistant to interference in the surrounding environment due to its reliance on carrier sense multiple access with collision avoidance (CSMA/CA).

\subsection{WiFi Sensing}

In WiFi sensing, it is common to first take the magnitude of the acquired CSI before using it for sensing \cite{Ma2019wifi}.\footnote{Instead of taking the magnitude, dividing between elements of a MIMO channel on a per-subcarrier basis is also a promising approach \cite{Zeng2021PCOM}. This operation also cancels out the absolute delay.}
On the other hand, using the impulse response, which is the Fourier pair of CSI, for WiFi sensing is not common, despite it revealing various details of the environment and targets of interest, as demonstrated in numerous studies on radar and channel sounding.\footnote{\cite{Xie2018} proposed to combine CSI observed across multiple adjacent channels to obtain a precise impulse response; however, it is not delay-synchronous.}
This is primary due to the poor quality of legacy COTS WiFi devices, which are subject to in-phase/quadrature imbalance among others.
However, the newer generation of WiFi devices offers much improved transceiver quality, allowing us to leverage the phase information of the transfer function for sensing, in addition to the envelope.



\section{SoundiFi: Delay-Synchronous Channel Sounding Framework}
\label{sec:sounding_framework}

\subsection{Delay in Observed Impulse Response}

The principle of delay-synchronous channel sounding is as follows.
If the Tx and Rx devices, the latter referred to as the {\it target}, are delay-synchronized as shown in Fig.~\ref{fig:1}\subref{fig:principle}, the delay $\tau$ of the observed impulse response at time $t$, denoted by $h(\tau,t)$, can be expressed relative to the transmission timing of the transmitted signal.

However, in the system considered here, 
the Tx and the target are not delay-synchronized.
Consequently, the transmission timing 
is unknown.
%
Additionally, at every CSI acquisition, delay zero on the target side can be considered the start time of OFDM symbol reception, where the reception timing has a degree of freedom within the cyclic prefix.
In this case, the reception start time relative to the Tx transmission time varies with each reception, that is, with each CSI acquisition.
Thus, a consistent reference for delay across acquisitions cannot be established.

\subsection{Antenna Remoting and Reference Antenna}

\begin{figure}[t]
\subfloat[][Basis.]{ 
\includegraphics[width=.3\linewidth,trim=0 2mm 0 3mm,clip]{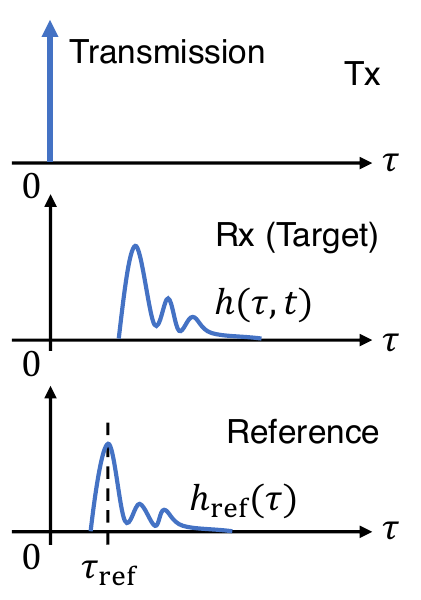}
 \label{fig:principle}
} \hfil
\subfloat[][COTS device setup.]{ 
 \includegraphics[width=.65\linewidth]{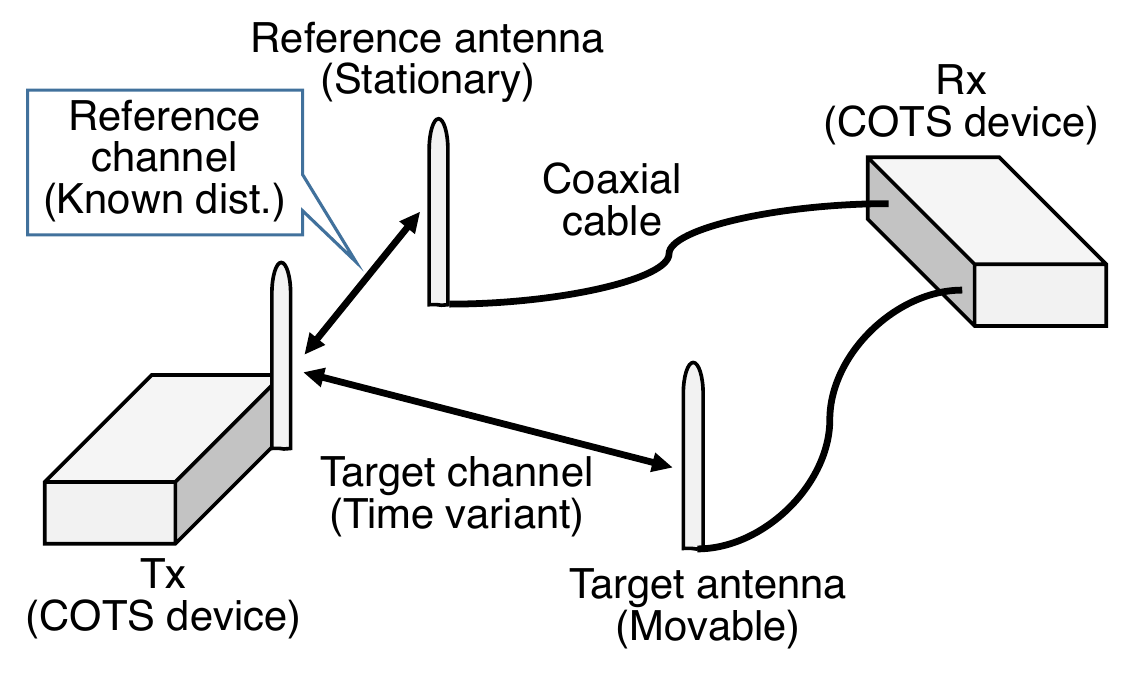}
 \label{fig:system_model}
}
\caption{SoundiFi: Delay-synchronous channel sounding framework.}
\label{fig:1}
\end{figure}



We propose SoundiFi, a framework designed to enable delay-synchronous channel sounding using COTS WiFi devices.
The key idea behind SoundiFi is to dedicate one of the MIMO channels as a reference, with a known over-the-air propagation distance, denoted as $d_\mathrm{ref}$.
As shown in Fig.~\ref{fig:1}\subref{fig:system_model}, the Rx antennas are connected via coaxial cables with identical lengths, with one Rx antenna fixed near the Tx while ensuring a line of sight with the Tx.
This antenna is referred to as the \textit{reference} antenna, while the other Rx antenna is referred to as the \textit{target} antenna.
The channels between the Tx and the reference or target antennas are called the reference and target channels, respectively.
For accurate calibration, it is desirable for the reference channel to remain stable throughout the series of acquisitions.\footnote{Reference CSI-based calibration was also proposed in \cite{Keerativoranan2018}, using a coaxial cable-connected channel between the Tx and Rx as the reference.}


According to the propagation distance of the reference channel, $d_\mathrm{ref}$, the delay of the first arrival relative to the transmitted signal is estimated as $\tau_\mathrm{ref}= d_\mathrm{ref}/c$, where $c$ represents the speed of light.
Let the impulse response of the reference channel be denoted by $h_\mathrm{ref}(\tau)$.
Note that the impulse response and CSI in this paper are represented in the equivalent baseband form.
As shown in Fig.~\ref{fig:1}\subref{fig:principle},
$\tau_\mathrm{ref}$ corresponds to the first signal of $\lvert h_\mathrm{ref}(\tau) \rvert$.
To represent this, we introduce the operator $\argfirstpeak$, which denotes the $\tau$ value corresponding to the first signal, i.e.,
\begin{align}
 \argfirstpeak_{\tau} \, \lvert h_\mathrm{ref}(\tau) \rvert &= \tau_\mathrm{ref}.
 \label{eq:peak_timing}
\end{align}
We refer to the operation of calibrating the observed reference and target impulse responses to satisfy (\ref{eq:peak_timing}) as delay calibration.
For delay calibration, the coaxial cables must be of the same length.
Additionally, we assume that 
variations in the lengths of internal wiring within the device are negligible.
The specific calibration procedure will be described in the next section.

\subsection{Procedure for Estimating Impulse Response from CSI}

Let the acquired reference and target CSI of the $k$th subcarrier at observation timing $t$ be denoted by $ \tilde{H}_\mathrm{ref}(k \varDelta f, t) $ and $ \tilde{H}(k \varDelta f, t)$, respectively, where $\varDelta f$ represents subcarrier spacing.
Note that in this paper, as in many WiFi sensing studies, the term CSI is used to mean the transfer function.
The average power spectrum of the reference CSI was adjusted to match the path loss between the Tx and reference, assuming free-space propagation.
Additionally, the average power spectrum of the target CSI was adjusted using the difference in received signal strength between the reference and target.


Let the impulse responses corresponding to these CSIs be denoted by $\tilde{h}_\mathrm{ref}(\tau,t)$ and $\tilde{h}(\tau,t)$, respectively.
Note that $\tilde{h}_\mathrm{ref}(\tau,t)$ represents the impulse response before the delay calibration, and thus, unlike ${h}_\mathrm{ref}(\tau)$, is time-variant with respect to delay across acquisitions though the shape of impulse responses remains nearly unchanged.

These impulse responses are obtained through the inverse discrete Fourier transform (IDFT) with respect to 
$k$, i.e., 
\begin{align}
 \tilde{h}_\mathrm{ref}(n \varDelta \tau , t) &= \mathsf{IDFT} [\tilde{H}_\mathrm{ref}(k \varDelta f, t)], \\
 \tilde{h}(n \varDelta \tau, t) &= \mathsf{IDFT} [\tilde{H}(k \varDelta f, t)],
\end{align}
where $\varDelta \tau$ represents the delay bin width, with the need to account for the oversampling described later.

The CSI acquisition effectively limits the bandwidth to the product of the subcarrier spacing and the number of subcarriers, 158.2\,MHz. As a result, directly applying the above IDFT leads to significant delay-domain sidelobes. 
To suppress sidelobes, a Blackman window was applied in the frequency domain before performing the IDFT.
The reduction of total channel power in the impulse response due to windowing was compensated so that the peak value matches the one obtained without the windowing.

Additionally, oversampling with a factor of $\kappa = \text{8}$ in the delay domain is achieved by extending the bandwidth $B=\text{160\,MHz}$ to $\kappa B$ through zero padding in the frequency domain. The resulting delay bin width becomes $\varDelta \tau = {1}/{\kappa B} = \text{0.78\,ns}$, corresponding to 0.23\,m wave propagation distance. 
Note that the delay resolution of the impulse response is about $\text{1/158.2\,MHz} = \text{6.3\,ns}$, corresponding to 1.9\,m. 

%

\subsection{Delay Calibration and Phase Normalization}
\label{ssec:delay_calibration}

The delay calibration of the target impulse response $\tilde{h}(\tau, t)$ using the reference $\tilde{h}_\mathrm{ref}(\tau, t)$ to obtain $h(\tau, t)$ is performed for every $t$.
First, the first signal of the reference impulse response is detected, and the phase at that signal is determined, i.e.,
\begin{align}
 \tau_t^\star &= \argfirstpeak_\tau \, \lvert \tilde{h}_\mathrm{ref}(\tau, t) \rvert,\ 
 \theta_t^\star = \arg \tilde{h}_\mathrm{ref}(\tau_t^\star, t ).
\end{align}

Then, the delay and phase of the reference and target impulse responses are shifted so that they satisfy (\ref{eq:peak_timing}), i.e.,
\begin{align}
 h_\mathrm{ref}(\tau) &= \tilde{h}_\mathrm{ref}(\tau -(\tau_\mathrm{ref} - \tau_t^\star), t)\, \ee^{-\jj \theta_t^\star}, \\ 
 h(\tau, t) &= \tilde{h}(\tau - (\tau_\mathrm{ref} -\tau_t^\star), t)\, \ee^{-\jj \theta_t^\star}. 
\end{align}
In this operation, the delay shift by $\tau_\mathrm{ref}-\tau_t^\star$ is referred to as delay calibration, while the phase shift by $\ee^{-\jj \theta_t^\star}$ is referred to as phase normalization.
Through phase normalization, the phase of the target is represented relative to the phase at the first signal of the reference impulse response.
As a result, the phase of the target becomes continuous across acquisitions, which is useful for averaging out the noise.


%

\section{Experimental Evaluation}
\label{sec:experimental}

To validate the feasibility of SoundiFi, we conducted evaluations using CSI observations obtained from COTS devices.

\subsection{Experimental Equipment}


As the Tx and Rx, we used two ASUS NUC 13 Rugged-Tall, each equipped with an Intel AX210 WiFi card and running PicoScenes 2024.1003.0247 \cite{Jiang2022IOT} for CSI acquisitions.
Both the Tx and Rx were equipped with two removable antennas with RP-SMA connectors, each of which could be connected via RP-SMA coaxial cables.
The use of two antennas allows for acquisition of the $2 \times 2$ MIMO channel.

Note that, in the subsequent evaluations, only the CSI of the $1 \times 2$ channel between one Tx antenna and the Rx antennas will be used as the reference and target channels.
In other words, although channels corresponding to the other Tx antenna can also be observed and used as reference and target channels, they are excluded from the following evaluation to keep the discussion concise.

As shown in Fig.~\ref{fig:1}\subref{fig:system_model}, two 5\,m RP-SMA coaxial cables connected the Rx and its antennas. 
The Rx ports of the IEEE 802.11ax device were covered with electromagnetic shielding fabric
to suppress possible spurious peaks due to electromagnetic leakage from any components in the device.

The packets transmitted by the Tx comply with IEEE 802.11ax, with a bandwidth of 160\,MHz, a center frequency of 5250\,MHz, and a subcarrier spacing of 78.125\,kHz.
There are 2025 subcarriers, indexed from $-$1012 to 1012.
The Tx sends packets at 8\,ms intervals, and the Rx acquires the CSI and received power for each of the two Rx antennas.

Regarding the packet transmission interval, it should be noted that, due to the random backoff in WiFi, there is uncertainty in the transmission timing, even when no other transmitters are present.
Specifically, when the minimum contention window is 15 and the slot time is 9\,$\mu$s, the transmission timing can vary by up to 135\,$\mu$s.
While PicoScenes can reduce the transmission interval to as small as 1\,ms, reducing the transmission interval increases the impact of this uncertainty.

\begin{figure}[t]
 \centering
 \includegraphics[width=\linewidth]{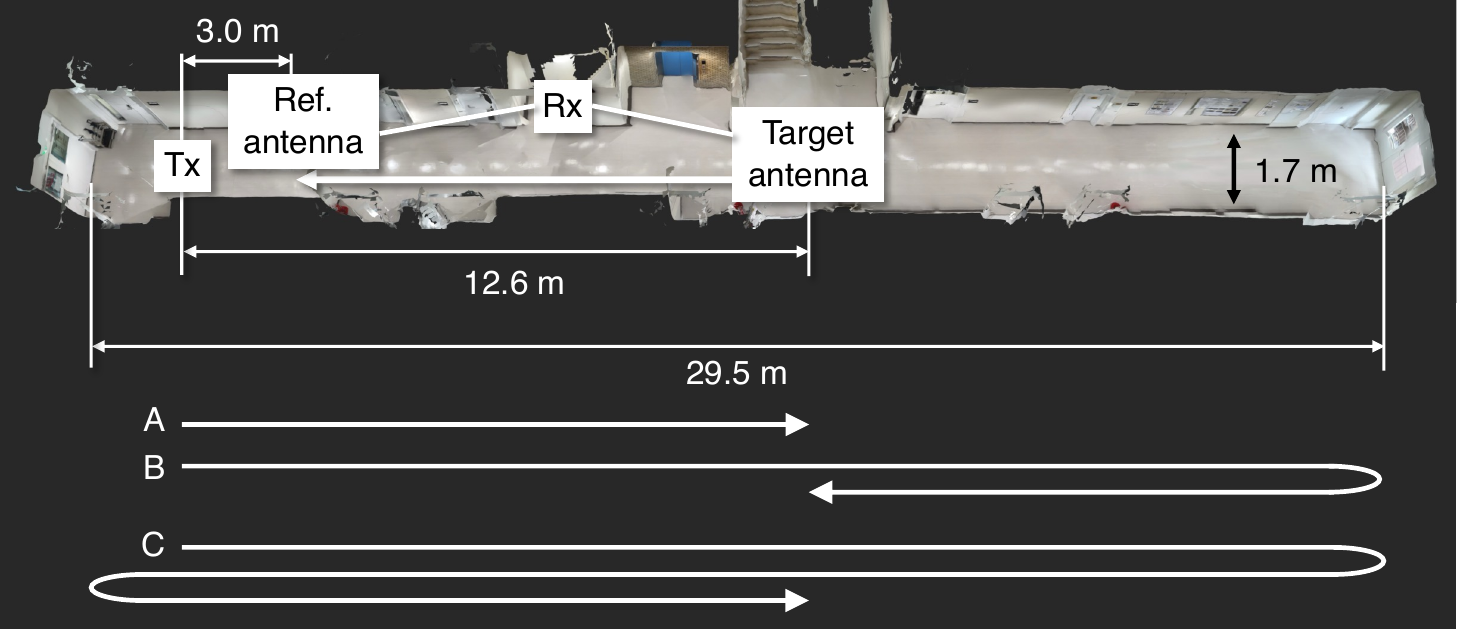}
 \caption{Sounding site. Paths A, B, and C are described in Section \ref{ssec:ir}.}
 \label{fig:hallway}
\end{figure}

\subsection{Sounding Site}

CSI observation was conducted in the corridor shown in Fig.~\ref{fig:hallway}, which is 29.5\,m in length, 2.5\,m in height, and 1.7\,m in width at its narrowest point, with rooms located on both the longitudinal sides.
The Tx and reference antenna were installed $d_\mathrm{ref}=\text{3.0\,m}$ apart at a height of 0.8\,m, while the target antenna, installed on a cart at the same height, was pushed by a person behind the cart to move it from a point 12.6\,m away from the Tx to a point near the reference antenna.
The Rx was fixed at the midpoint between the reference antenna and the initial position of the target antenna.
While moving the target antenna, an \textit{optical} line of sight with the Tx was always maintained.

\subsection{Preprocessing}

As part of the preprocessing, we used PicoScenes Python Toolbox to interpolate the CSI at unobserved pilot subcarriers. 
Additional procedures we performed are explained below.

The CSI of the DC subcarriers (indices $-$11, $-$10, \ldots, 11) was not observed because no signal was transmitted on those subcarriers, as specified by IEEE 802.11ax, and both their magnitude and phase were linearly interpolated using those of neighboring subcarriers.\footnote{PicoScenes Python Toolbox outputs linearly interpolated phases, which can  be discontinuous and introduces  noise into the impulse response.}
Note that, since the phase has an uncertainty of multiples of $2\pi$, the unwrapped phase at subcarrier index 12 was first determined by linearly extrapolating the phases at the DC subcarriers using the unwrapped phases of subcarriers with indices $-$21 to $-$12.
Then, the phases at DC subcarriers were determined through linear regression using the phases at subcarrier indices $\pm$12 to $\pm$12.

In the raw observations of the CSI, the values corresponding to the two Rx antennas were occasionally swapped during acquisition.
This was corrected based on the distance between the magnitude sequences of the current and previous acquisitions.

Additionally, attenuation likely due to device characteristics rather than the wave propagation was observed at subcarriers $\pm$766 to $\pm$770.
Subsequently, the measured CSI magnitudes from those subcarriers were discarded, and their values were linearly interpolated using the magnitudes of neighboring subcarriers.
Furthermore, the time-averaged power was not uniform across subcarriers, with the average power at the edges of the spectrum being smaller than at the center.
Therefore, we normalized the power to make it uniform across all subcarriers.

Although not strictly preprocessing, it was observed that during phase normalization, the phase of the target CSI occasionally undergoes an abrupt $\pi$ shift.
To address this, the impulse response for the next time step was chosen as the one with the smallest distance to the impulse response derived from the previous acquisition, between the current impulse response and its $\pi$-shifted version.

\subsection{Reference}

\begin{figure}[t]
 \vspace{-12pt}
 \centering
 \subfloat[][Before calibration.]{
 \includegraphics[height=.595\linewidth] {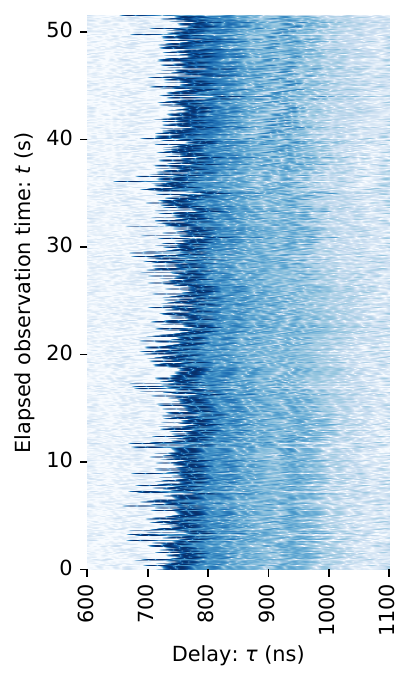}
 \label{fig:00_noroll}
 } 
 \subfloat[][After calibration.]{
 \includegraphics[height=.6\linewidth]{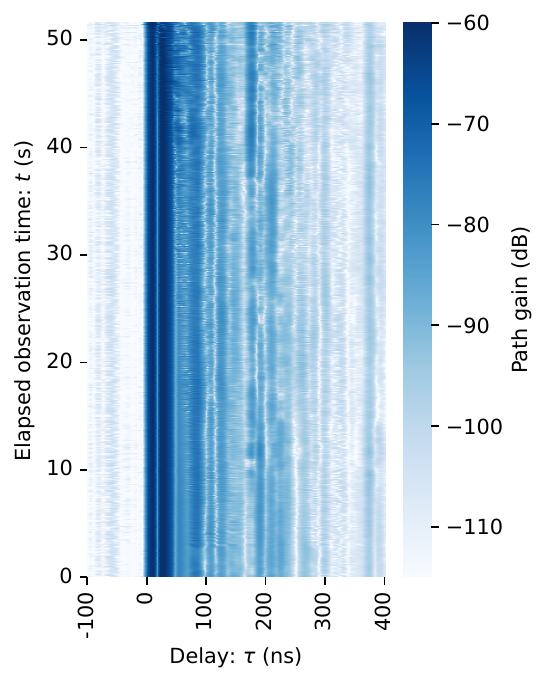}
 \label{fig:00_roll}
 } 
 \caption{Time series of the reference impulse response.}
 \label{fig:00}
\end{figure}

Fig.~\ref{fig:00} shows the time series of the reference impulse response.
More specifically, Figs.~\ref{fig:00}\subref{fig:00_noroll} and \ref{fig:00}\subref{fig:00_roll} correspond to those before and after the calibration, where the reference first signal is aligned to delay $\tau_\mathrm{ref}=d_\mathrm{ref}/c=\text{10\,ns}$.
Fig.~\ref{fig:00}\subref{fig:00_noroll} shows that the first signals, which should be at the identical delay, fluctuate between acquisitions.
This is because, as mentioned in Section~\ref{sec:introduction}, the OFDM symbol reception timing has a degree of freedom within the cyclic prefix, causing the timing of the first signal to fluctuate depending on when the symbol reception is started within that range.

\begin{figure}[t]
 \centering
 \includegraphics[width=.8\linewidth]{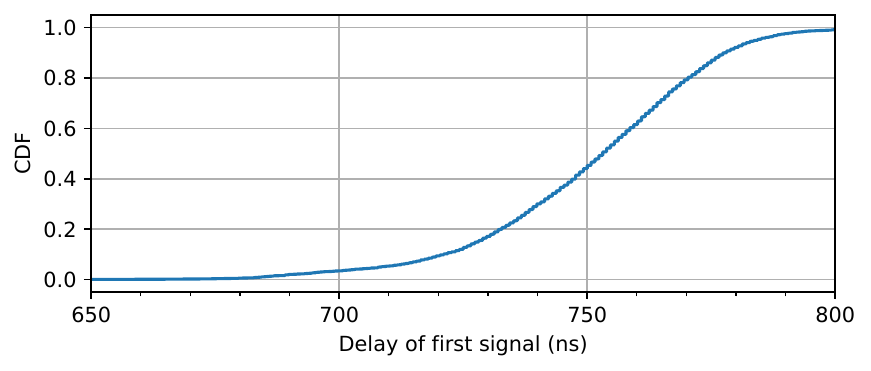}
 \caption{Empirical CDF of first signal delay before calibration.}
 \label{fig:CDF_delay}
\end{figure}

Fig.~\ref{fig:CDF_delay} shows the empirical cumulative distribution function (CDF) of the delay of the first signal in Fig.~\ref{fig:00}\subref{fig:00_noroll}.
The delay is observed to be nearly uniformly distributed in the range from 680\,ns to 800\,ns.
Note that the cyclic prefix in the present transmission is set to 3200\,ns, so such variation falls well within a feasible range.

On the other hand, Fig.~\ref{fig:00}\subref{fig:00_roll} shows that, after delay calibration, the impulse response remains stable in terms of delay across observations.
Therefore, the reference can be considered useful for calibrating the delay of the target impulse response.

\subsection{Spectrum}

\begin{figure}[t]
 \centering
 \subfloat[][Path gain.]{
 \includegraphics[width=0.8\linewidth]{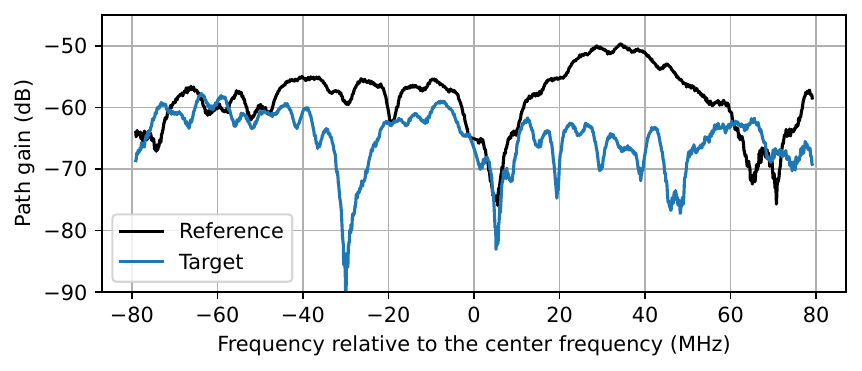}
 \label{fig:H1x2_amp}
 } \\
 \subfloat[][Unwrapped phase.]{
 \includegraphics[width=0.8\linewidth]{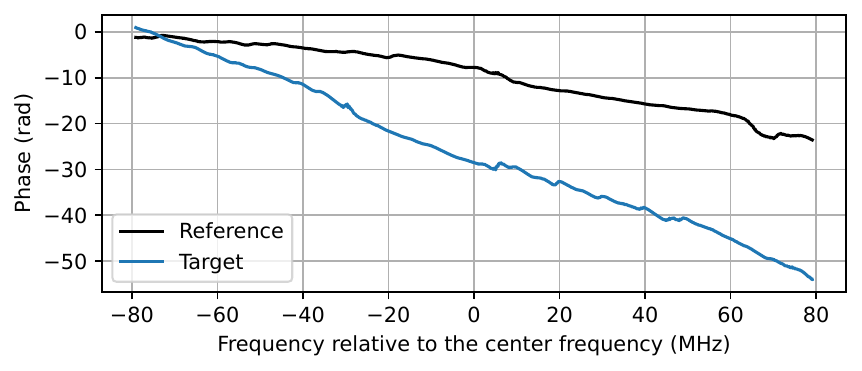}
 \label{fig:H1x2_phase}
 }
 \caption{CSI at experiment start, $t=\text{0\,s}$. }
 \label{fig:H1x2}
\end{figure}

As an example of acquired CSI, Fig.~\ref{fig:H1x2} shows a snapshot at $t=\text{0\,s}$, with the target antenna positioned farthest from the Tx.
Fig.~\ref{fig:H1x2}\subref{fig:H1x2_phase} shows the unwrapped phase after delay calibration. 
The steeper negative slope of the target CSI compared to the reference indicates a delay in the dominant signal of the target relative to the reference.
This is because the slope in the figure corresponds to the delay of the dominant signal.

\subsection{Impulse Response}
\label{ssec:ir}

\begin{figure}[t]
 \centering
 \subfloat[][Target impulse response at $t=\text{0\,s}$.]{
 \includegraphics[width=0.8\linewidth]{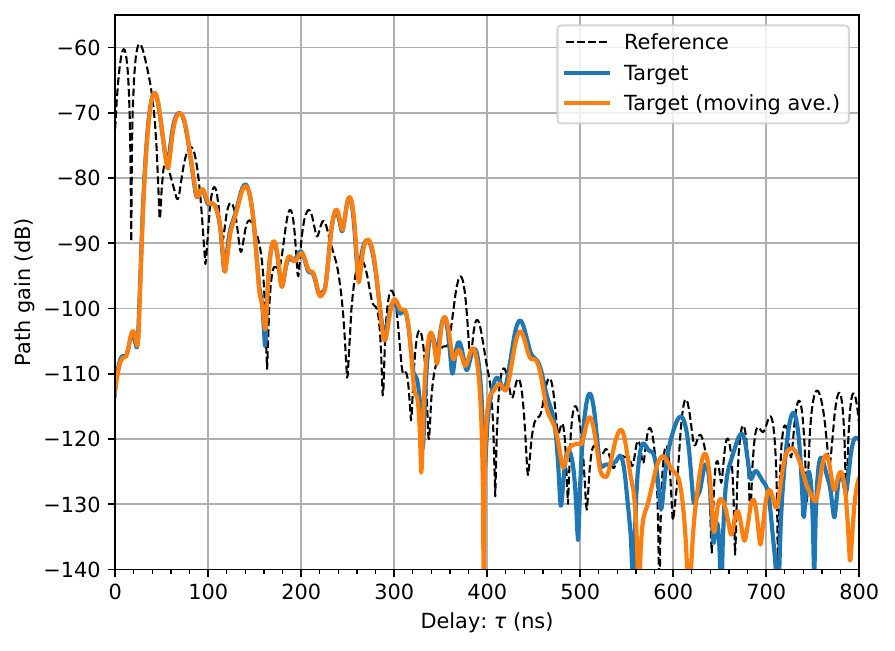}
 \label{fig:IR_DFT_log}
 } \\
 \subfloat[][Time series of target impulse response.]{
 \includegraphics[width=.9\linewidth]{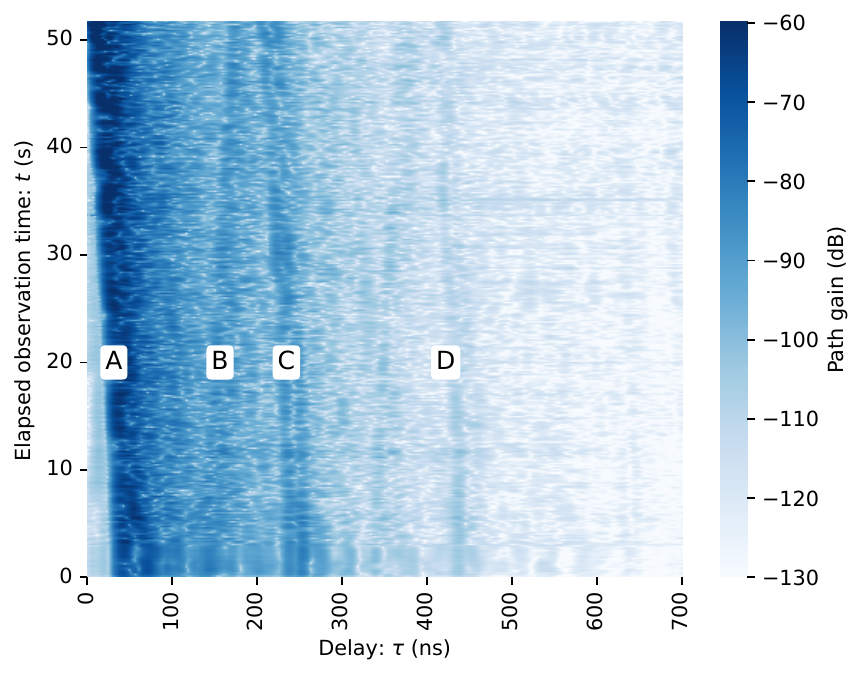}
 \label{fig:peaks}
 }
 \caption{Target impulse response, $h(\tau, t)$. A moving average was applied in the time domain, with a window size of 11.}
 \label{fig:IR}
\end{figure}


Fig.~\ref{fig:IR}\subref{fig:IR_DFT_log} shows the target impulse response, and Fig.~\ref{fig:IR}\subref{fig:peaks} shows its corresponding time series.
It can be confirmed that the horizontal axis is adjusted such that the first signal of the reference occurs at $\tau_\mathrm{ref}= \text{10\,ns}$.

The noise level is $-\text{115\,dB}$, and by applying a moving average with a window size of 11 along the time axis to the delay-calibrated CSI, the noise level is further reduced to $-\text{122\,dB}$.
It is possible to recognize peak signals as a propagation path if they are well above this noise level.
It should be noted that noise reduction through the moving average is possible due to the phase continuity across acquisitions achieved through the phase normalization introduced in Section~\ref{ssec:delay_calibration}.

There are three noticeable peak signals at $\tau=\text{40\,ns}$, 240\,ns, and 440\,ns at $t=0$, labeled A, C, and D in Fig.~\ref{fig:IR}\subref{fig:peaks},
indicating the successful observation of long-delayed multipaths with at least 132\,m propagation distance.
The power of these signals decreases monotonically in $\tau$ as $t$ increases.
This result can be understood by the fact that the target is moving steadily closer to the Tx, reducing the propagation distance.
The delay difference between signal A and signal C/D is 200\,ns/400\,ns, corresponding to a propagation path length of 60\,m/120\,m that would represent one or two round trips along the corridor, as shown in Fig.~\ref{fig:hallway}.
On the other hand, signal B shows an increasing path length from the Tx as $t$ increases, which corresponds to the wave reflecting once from the right end of the corridor as illustrated in Fig.~\ref{fig:hallway}.


\vfill

\section{Conclusion}

We proposed and experimentally demonstrated SoundiFi, a delay-synchronous channel sounding framework using COTS WiFi devices.
The key aspects of SoundiFi include antenna remoting via coaxial cables and the use of a reference antenna with known distance to the Tx, enabling accurate delay calibration for the target channel. 

We believe that SoundiFi facilitates accessibility to delay-synchronous channel sounding, offering a cost-effective and license-exempt alternative to traditional methods. 
Additionally, while SoundiFi cannot be directly applied to WiFi sensing scenarios using everyday WiFi devices, it enables more accurate delay and phase measurements, which are critical for applications such as object tracking in dynamic environments. This stands in contrast to conventional WiFi sensing, which does not fully leverage delay and phase information.



To extend SoundiFi to multiple target channels, multiple receivers calibrated by their reference distances or a device with more than two antennas could be used.
Alternatively, multiple receivers could be connected via a splitter from a common reference antenna.



\vfill

\section*{Acknowledgment}

This work was supported in part by JSPS KAKENHI Grant Numbers JP23K24831 and JP23K26109.
We would like to thank Dr. Akihito Taya and Dr. Takayuki Nishio for their advice on CSI acquisition.


\vfill 

\bibliographystyle{IEEEtran}
\bibliography{IEEEabrv,otherabrv,kyamamot-ct,library2}

\end{document}